\documentclass[
    ,final            
  ]
  {aipproc}

\layoutstyle{6x9}

\begin{document}

\title{Ultraviolet Dust Grain Properties in Starburst Galaxies:
Evidence from Radiative Transfer Modeling and Local Group Extinction Curves}

\author{Karl D.\ Gordon}{
  address={Steward Observatory, University of Arizona, Tucson, AZ
85721, USA \\ email: kgordon@as.arizona.edu}
}

\begin{abstract}
This paper summarizes the evidence of the ultraviolet properties of
dust grains found in starburst galaxies.  Observations of starburst
galaxies clearly show that the 2175~\AA\ feature is weak or absent.
This can be the result of radiative transfer effects (mixing
the dust and stars) or due to dust grains which do not have
this feature.  Spherical DIRTY radiative transfer models imply that it
is not radiative transfer effects, but other radiative transfer models
with disk/bulge geometries have found cases where it could be
radiative transfer effects.  Recent work on the extinction curves in
the Magellanic Clouds and Milky Way has revealed that the traditional
explanation of low metallicity for the absence of the 2175~\AA\
feature in the Small Magellanic Cloud is likely incorrect.  The SMC
has one sightline with a 2175~\AA\ feature and the Milky Way has
sightlines without this feature.  In addition, where the 2175~\AA\
feature is found to be weak or absent in both Magellanic Clouds and
the Milky Way, there is evidence for recent star formation.  Taking
the sum of the radiative transfer modeling of starburst galaxies and
the behavior of Local Group extinction curves, it is likely that the
dust grains in starburst galaxies intrinsically lack the 2175~\AA\
feature.
\end{abstract}

\maketitle


\section{Introduction}

There are two main reasons to study the dust in galaxies where single
stars are unresolvable with current telescopes.

1) To directly study the dust itself as galaxies provide a much wider
range of environments than found in the Local Group.  These different
environments can be characterized by more extreme values of radiation
field density, radiative field hardness, shock frequency, and initial
conditions (eg., metallicity).  Such different environments can
significantly impact the formation and destruction of dust grains.

2) To more accurately account for the effects of dust on observations
of the stars and gas present in galaxies.  This accounting allows for
the study of stars and gas in galaxies to be carried out more
accurately. 

In order to study dust in such galaxies, a model of how the mixing of
dust with stars and gas affects to the observations of a galaxy is
required.  One such model is the DIRTY dust radiative transfer model
\citep{Gordon01, Misselt01} which uses Monte Carlo techniques to
compute the radiative transfer of photons through dust and
self-consistently accounts for the dust re-emission in the infrared
including equilibrium and non-equilibrium emission.

This paper concentrates on the ultraviolet dust grain properties in
starburst galaxies determined using the DIRTY model and implied from
extinction curves measured in the Local Group.  The study of dust in
galaxies is much larger than this, especially in light of the wealth
of information available in the infrared where the dust grain emission
dominated the spectral energy distribution of a galaxy.  The
connection between ultraviolet and infrared dust grain properties is
not direct, but determined using dust grain models
\citep{Weingartner01, Clayton03a, Zubko04} which have
their own problems.  For example, I have found that using empirically
determined dust scattering properties in the ultraviolet
\citep{Gordon04} is crucial to reproducing the ultraviolet colors of
starburst galaxies.

\section{Starburst Galaxies}

Starburst galaxies provide ideal environments to probe dust grain
properties in extreme environments as they are intrinsically bright in
the ultraviolet where dust grain properties show large variation
\citep{Cardelli89,Fitzpatrick90,Gordon03}.  The environments
probed by starbursts range from very metal poor (eg., I Zw 18) to very
metal rich and most starbursts are likely
characterized by high and hard radiation fields and elevated shock
frequencies.  Starbursts probe the more active
environments of those possible in galaxies in general.

The study of the type of dust found in starburst galaxies really got
going with the work of \citet{Calzetti94} and \citet{Calzetti97}.  In
these studies, a variant of the standard pair method was used to
derive an empirical attenuation curve appropriate for ultraviolet
bright starburst galaxies.  This empirical curve clearly lacked the
strong 2175~\AA\ absorption feature seen in almost all known dust
extinction curves.  In addition, this curve was grayer than most dust
extinction curves.  The question was then: Is the lack of this
2175~\AA\ feature and the grayer curve due to different dust grain
properties or radiative transfer effects?  At the time, only a small
number of sightlines in the Small Magellanic Cloud (SMC) were known to
lack this feature \citep{Prevot84}.  This lack was attributed to the
low metallicity of the SMC with the higher metallicities of the Large
Magellanic Cloud and the Milky Way resulting in the ubiquity of the
2175~\AA\ feature in their dust \citep{Cardelli89,Fitzpatrick90}.
But, the starburst galaxies studied by \citet{Calzetti94} had
metallicities spanning the known range from lower than the SMC to
higher than the Milky Way.  Radiative transfer effects were seen as
the probable cause of the lack of the 2175~\AA\ feature.  The
complicated mixing of the dust, gas, and many stars in unresolved
observations of galaxies implies that radiative transfer effects are
important.  Such mixing means that the stars in the observing beam
have different dust columns and significant scattered flux is included
in any measurement.  As a result, the attenuation curve for a galaxy
is not dependent only on dust grain properties, like that for the
extinction curve towards a single
resolved star in the Local Group, but is also dependent on the
geometry of the dust, gas, and stars.

Motivated by the starburst results and improvements in the
DIRTY dust radiative transfer model to include nonhomogeneous dust
distributions \citep{Witt96}, \citet{Gordon97} investigated the
importance of radiative transfer effects on determining the shape of
the empirical starburst attenuation curve.  After investigating a number
of dust/star geometries, they found that the lack of the 2175~\AA\
feature could not be explained by radiative transfer effects.  The
only way to remove the 2175~\AA\ feature from the attenuation curve
was for the dust grains themselves to lack the feature.  This result
coupled with the large range in metallicities in the starburst
galaxies implied that it was the starburst environment which was
responsible for the lack of the 2175~\AA\ feature.  The dust grains
were either newly formed without the feature or this feature was
destroyed by the hard radiation field and elevated shock frequency in
starbursts.  Using the same or similar analysis, this result was
extended to higher redshifts ($z \approx 3$) by \citet{Gordon99} and
\citet{Vijh03} where it was also found that most high-redshift (and
likely starburst) galaxies lack a 2175~\AA\ bump.  A detailed analysis
of the \citet{Calzetti97} attenuation curve by \citet{Witt00} found
that a clumpy SHELL geometry with SMC dust, $\tau_V \sim 1.5$ DIRTY
model best fit both 
the grayness of the curve and the lack of the 2175~\AA\ feature.

Since the work utilizing simple spherical DIRTY models, other studies
have found ways to produce attenuation curves with weak or nearly
absent 2175~\AA\ features while using dust grains with Milky Way-like
dust (i.e., a strong 2175~\AA\ feature).  Using the GRASIL code,
\citet{Granato01} found that differentially 
embedding sources of different spectral types (basically by stellar
lifetime) in molecular clouds in a disk/bulge geometry with clumpy
structure and Milky Way-like dust resulted in attenuation curves with
weak or absent 2175~\AA\ features.  \citet{Fischera03} studied the
attenuation curves produced by a foreground screen with a log-normal
density distribution of Milky Way-like dust and was able to reproduce
the grayness of the \citet{Calzetti97} curve, but was only able to
reduce the strength of the 2175~\AA\ feature.
Finally, \citet{Pierini04} used DIRTY models of disk/bulge galaxies
and found that for highly inclined, weak bulge galaxies the 2175~\AA\
feature was very weak in the global attenuation curves.  

As a result of these studies \citep{Granato01,Fischera03,Pierini04},
it is clear that there are dust/star geometries which can reproduce
most, if not all, of the characteristics of the \citet{Calzetti97}
starburst curve.  But it is not clear that the necessary geometries
(disk/bulge or foreground screen) are appropriate for the starburst
galaxies which are usually seen as nuclear dominated or have highly
irregular morphologies.

\section{Local Group Extinction Curves}

An important part of the type of dust we expect to find in galaxies is
driven by our detailed knowledge of dust in the Local Group.  One of
the most powerful ways to study the properties of dust is to determine
ultraviolet through near-infrared extinction curves towards single
stars.  As the dust measured by observing a star is distributed in a
foreground screen, the derived extinction curve is only dependent on
the properties of the dust grains (size, composition, and shape).
With current telescopes and instruments, this work is limited to Local
Group galaxies.

Prior to the work on starburst galaxies, it was thought that most
galaxies would have dust with a 2175~\AA\ feature because all known
sightlines in the Milky Way \citep{Cardelli89,Fitzpatrick90} and the
Large Magellanic Cloud \citep{Clayton85,Fitzpatrick86} had this
feature.  Only in the 
Small Magellanic Cloud was the 2175~\AA\ feature absent
\citep{Prevot84} and this was understood to be due to the low
metallicity of this galaxy.  The lack of the 2175~\AA\ feature in
starburst galaxies with metallicities like those found in the Milky
Way and LMC questioned the explanation for the SMC lacking the
2175~\AA\ feature.  In fact, the variation in the LMC extinction
curves between those near and far from the 30~Dor star forming region
pointed to a different explanation \citep{Fitzpatrick86}.  It was
found that the extinction curves near 30~Dor had weaker 2175~\AA\
features and somewhat steeper far-ultraviolet extinctions than those
found in the rest of the LMC.  

The existence of two explanations (low metallicity versus nearby
active star formation) for the weakness or lack of a 2175~\AA\ feature
motivated new work on extinction curves in the Local Group.

In the SMC, \citet{Gordon98} searched the IUE archives and found it
was possible to generate accurate extinction curves for only four
sightlines.  Three of the four curves were basically linear with
$1/\lambda$ and had no detectable 2175~\AA\ feature.  The fourth curve
was much more like that found in the Milky Way with a 2175~\AA\
feature and weaker far-ultraviolet extinction compared to the rest of
the SMC curves.  This Milky Way-like curve was first published in
\citet{Lequeux82} but dismissed by \citet{Prevot84} as anomalous.
\citet{Gordon98} found that this Milky Way-like curve was located in a
more quiescent region than the other three curves which were located in
the star forming bar of the SMC.  One additional sightline was added
to the sample of SMC extinction curves by \citet{Gordon03} using new
HST/STIS observations.  This new sightline was in the star forming bar
and was very similar to the previously known extinction curves in this
same region.  Thus, the spatial variations in the five known
extinction curves in the SMC are consistent with the explanation that
the lack of the 2175~\AA\ feature is due to nearby active star
formation.

In the LMC, \citet{Misselt99} searched the IUE archives and was able
to construct 19 accurate extinction curves from the available data.
Like previous studies, this work found significant differences between
those sightlines near 30~Dor and those in the rest of the LMC.
\citet{Misselt99} found that this difference was caused by the
sightlines associated with the supergiant shell LMC 2 near 30~Dor, not
30~Dor itself.  Thus, they found very significant differences
(especially in their 2175~\AA\ feature strengths) when the
19 sightlines were grouped between those associated with the LMC 2
shell and those in the rest of the LMC.  Like the SMC, the spatial
variation of the LMC extinction curves is consistent with active star
formation being the cause.

In the Milky Way, the most comprehensive extinction curve study to
date has been the recent work of \citet{Valencic04}.  Like the studies
in the SMC and LMC \citep{Gordon98,Misselt99}, the IUE archive was
searched and a total of 417 extinction curves created.  The
overwhelming majority of these curves (93\%) were well characterized by
the \citet{Cardelli89} $R_V$ dependent relationship with only 4 curves
showing systematic deviation from this relationship.  HD~204827 is one
of these 4 curves and a detailed study of this sightline
\citep{Valencic03} found that after subtracting a well measured
foreground component, the remaining extinction curve was equivalent
within the uncertainties to the extinction curves found in the SMC
star forming bar (i.e., linear with $1/\lambda$ and lacking a 2175~\AA\
bump).  The local environment of HD~204827 has evidence for a recent
supernova shock.  In a study of 30 low reddening, long sightlines in the
Milky Way, \citet{Clayton00} found that a subsample which had
extinction curves like that found in the LMC 2 shell sample
\citep{Misselt99}.  This subsample were all in the same region of the
Milky Way and displayed N(Ca II)/N(Na I) ratios and velocities
indicating recent dust destruction.  Like the work in Magellanic
Clouds, new studies in the Milky Way have revealed evidence that
processing of dust can cause the weakening or disappearance of the
2175~\AA\ feature and strengthening of the far-ultraviolet
extinction.

Examining the sum of work on extinction curves in the Milky Way and
Magellanic Clouds, it is clear that there is a continuum of dust
extinction curves extending from those represented by the
\citet{Cardelli89} relationship with strong 2175~\AA\ features to
those like those found in the SMC bar with no detectable 2175~\AA\
feature \citep{Gordon03}.  This implies that it is more accurate to
describe the \citet{Cardelli89} relationship as referring to dust in
quiescent environments and SMC bar extinction curves as referring to
dust in much more active environments.  But even this picture of dust
ultraviolet extinction curve variations is incomplete as dust in
molecular clouds has not been well measured.  Only the Taurus
molecular cloud has been probed in the ultraviolet through the
HD~29647 and HD~283809 sightlines \citep{Clayton03b}.  By subtracting a
foreground extinction curve measurement from these two sightlines,
\citet{Whittet04} found that the 2175~\AA\ bump disappeared.  This is
like the dust found in the SMC bar, but the far-ultraviolet extinction
in this dust was much weaker.  This is good evidence that the true
range of dust extinction curves likely includes another extreme, that
found in molecular clouds.


\section{Discussion}

In this paper, I have attempted to summarize work I have been involved
in relating to the dust found in starburst galaxies and Local Group
extinction curve work motivated in part by the starburst galaxy
results.  Combining the results for radiative transfer in starburst
galaxies for what we now know is a more varied story of dust
extinction curves in the Milky Way and Magellanic Clouds, it is most
probable that the dust grains in starburst galaxies are truly lacking
the 2175~\AA\ feature and possesses strong far-ultraviolet extinctions.
It is very important to clearly state that this conclusion only
applies to the dust found in or near the starburst regions.  This is
the dust which is probed by the ultraviolet photons from the starburst
regions.  Given our knowledge of Local Group dust extinction curves,
it is very likely that dust far from sites of active star formation in
starburst galaxies would be characterized by Milky Way-like or
quiescent dust (i.e., \citet{Cardelli89} relationship).

The measurements of starburst galaxies in the ultraviolet necessarily
probe only the active regions of these galaxies; there are other
measurements which probe the type of dust found in more quiescent
regions of galaxies.  Studies of the colors of gravitational lens
systems \citep{Nadeau91,Falco99} have found \citet{Cardelli89} type
dust curves.  As gravitational lenses probe random sightlines in
galaxies, this is evidence that quiescent dust is like that seen in
our Galaxy.  These gravitational lens difference curves are not direct
measures of dust extinction as they probe the difference in two dust
extinction curves of unknown dust columns \citep{Wucknitz03,
Mcgough05}, but they can be used to determine the presence of 
the 2175~\AA\ feature.  Finally, measurements of Mg II absorbers also
probe random sightlines in galaxies and show evidence for the
2175~\AA\ feature \citep{Malhotra97,Wang04}.


\begin{theacknowledgments}
I would like thank the organizers of this conference for inviting me
to give a talk on my work.  The work described in the paper
encompasses a large body of work, which would not have been possible
but for the contributions of many of my collaborators.  I would like
to especially thank Adolf Witt who was my Monte Carlo radiative
transfer mentor and Geoff Clayton who was my extinction curve mentor.
\end{theacknowledgments}


\bibliographystyle{aipproc}   

\bibliography{sed2004_sbdust}

\begin{thebibliography}{36}
\expandafter\ifx\csname natexlab\endcsname\relax\def\natexlab#1{#1}\fi
\providecommand{\enquote}[1]{``#1''}
\expandafter\ifx\csname url\endcsname\relax
  \def\url#1{\texttt{#1}}\fi
\expandafter\ifx\csname urlprefix\endcsname\relax\def\urlprefix{URL }\fi

\bibitem[{Gordon} et~al.(2001)]{Gordon01}
{Gordon}, K.~D., {Misselt}, K.~A., {Witt}, A.~N., and {Clayton}, G.~C.,
  \emph{ApJ}, \textbf{551}, 269 (2001).

\bibitem[{Misselt} et~al.(2001)]{Misselt01}
{Misselt}, K.~A., {Gordon}, K.~D., {Clayton}, G.~C., and {Wolff}, M.~J.,
  \emph{ApJ}, \textbf{551}, 277 (2001).

\bibitem[{Weingartner} and {Draine}(2001)]{Weingartner01}
{Weingartner}, J.~C., and {Draine}, B.~T., \emph{ApJ}, \textbf{548}, 296
  (2001).

\bibitem[{Clayton} et~al.(2003{\natexlab{a}})]{Clayton03a}
{Clayton}, G.~C., {Wolff}, M.~J., {Sofia}, U.~J., {Gordon}, K.~D., and
  {Misselt}, K.~A., \emph{ApJ}, \textbf{588}, 871 (2003{\natexlab{a}}).

\bibitem[{Zubko} et~al.(2004)]{Zubko04}
{Zubko}, V., {Dwek}, E., and {Arendt}, R.~G., \emph{ApJS}, \textbf{152}, 211
  (2004).

\bibitem[{Gordon}(2004)]{Gordon04}
{Gordon}, K.~D., \enquote{{Interstellar Dust Scattering Properties},} in
  \emph{ASP Conf. Ser. 309: Astrophysics of Dust}, 2004, p.~77.

\bibitem[{Cardelli} et~al.(1989)]{Cardelli89}
{Cardelli}, J.~A., {Clayton}, G.~C., and {Mathis}, J.~S., \emph{ApJ},
  \textbf{345}, 245 (1989).

\bibitem[{Fitzpatrick} and {Massa}(1990)]{Fitzpatrick90}
{Fitzpatrick}, E.~L., and {Massa}, D., \emph{ApJS}, \textbf{72}, 163 (1990).

\bibitem[{Gordon} et~al.(2003)]{Gordon03}
{Gordon}, K.~D., {Clayton}, G.~C., {Misselt}, K.~A., {Landolt}, A.~U., and
  {Wolff}, M.~J., \emph{ApJ}, \textbf{594}, 279 (2003).

\bibitem[{Calzetti} et~al.(1994)]{Calzetti94}
{Calzetti}, D., {Kinney}, A.~L., and {Storchi-Bergmann}, T., \emph{ApJ},
  \textbf{429}, 582 (1994).

\bibitem[{Calzetti}(1997)]{Calzetti97}
{Calzetti}, D., \emph{AJ}, \textbf{113}, 162 (1997).

\bibitem[{Prevot} et~al.(1984)]{Prevot84}
{Prevot}, M.~L., {Lequeux}, J., {Prevot}, L., {Maurice}, E., and
  {Rocca-Volmerange}, B., \emph{A\&A}, \textbf{132}, 389 (1984).

\bibitem[{Witt} and {Gordon}(1996)]{Witt96}
{Witt}, A.~N., and {Gordon}, K.~D., \emph{ApJ}, \textbf{463}, 681 (1996).

\bibitem[{Gordon} et~al.(1997)]{Gordon97}
{Gordon}, K.~D., {Calzetti}, D., and {Witt}, A.~N., \emph{ApJ}, \textbf{487},
  625 (1997).

\bibitem[{Gordon} et~al.(1999)]{Gordon99}
{Gordon}, K.~D., {Smith}, T.~L., and {Clayton}, G.~C., \enquote{{Dust in High
  Redshift Starburst Galaxies},} in \emph{ASP Conf. Ser. 193: The Hy-Redshift
  Universe: Galaxy Formation and Evolution at High Redshift}, 1999, p. 517.

\bibitem[{Vijh} et~al.(2003)]{Vijh03}
{Vijh}, U.~P., {Witt}, A.~N., and {Gordon}, K.~D., \emph{ApJ}, \textbf{587},
  533 (2003).

\bibitem[{Witt} and {Gordon}(2000)]{Witt00}
{Witt}, A.~N., and {Gordon}, K.~D., \emph{ApJ}, \textbf{528}, 799 (2000).

\bibitem[{Granato} et~al.(2001)]{Granato01}
{Granato}, G.~L., {Silva}, L., {Bressan}, A., {Lacey}, C.~G., {Baugh}, C.~M.,
  {Cole}, S., and {Frenk}, C.~S., \emph{Astrophysics and Space Science
  Supplement}, \textbf{277}, 79 (2001).

\bibitem[{Fischera} et~al.(2003)]{Fischera03}
{Fischera}, J., {Dopita}, M.~A., and {Sutherland}, R.~S., \emph{ApJL},
  \textbf{599}, L21 (2003).

\bibitem[{Pierini} et~al.(2004)]{Pierini04}
{Pierini}, D., {Gordon}, K.~D., {Witt}, A.~N., and {Madsen}, G.~J., \emph{ApJ,
  in press} (2004).

\bibitem[{Clayton} and {Martin}(1985)]{Clayton85}
{Clayton}, G.~C., and {Martin}, P.~G., \emph{ApJ}, \textbf{288}, 558 (1985).

\bibitem[{Fitzpatrick}(1986)]{Fitzpatrick86}
{Fitzpatrick}, E.~L., \emph{AJ}, \textbf{92}, 1068 (1986).

\bibitem[{Gordon} and {Clayton}(1998)]{Gordon98}
{Gordon}, K.~D., and {Clayton}, G.~C., \emph{ApJ}, \textbf{500}, 816 (1998).

\bibitem[{Lequeux} et~al.(1982)]{Lequeux82}
{Lequeux}, J., {Maurice}, E., {Prevot-Burnichon}, M.-L., {Prevot}, L., and
  {Rocca-Volmerange}, B., \emph{A\&A}, \textbf{113}, L15 (1982).

\bibitem[{Misselt} et~al.(1999)]{Misselt99}
{Misselt}, K.~A., {Clayton}, G.~C., and {Gordon}, K.~D., \emph{ApJ},
  \textbf{515}, 128 (1999).

\bibitem[{Valencic} et~al.(2004)]{Valencic04}
{Valencic}, L.~A., {Clayton}, G.~C., and {Gordon}, K.~D., \emph{ApJ},
  \textbf{616}, 912 (2004).

\bibitem[{Valencic} et~al.(2003)]{Valencic03}
{Valencic}, L.~A., {Clayton}, G.~C., {Gordon}, K.~D., and {Smith}, T.~L.,
  \emph{ApJ}, \textbf{598}, 369 (2003).

\bibitem[{Clayton} et~al.(2000)]{Clayton00}
{Clayton}, G.~C., {Gordon}, K.~D., and {Wolff}, M.~J., \emph{ApJS},
  \textbf{129}, 147 (2000).

\bibitem[{Clayton} et~al.(2003{\natexlab{b}})]{Clayton03b}
{Clayton}, G.~C., {Gordon}, K.~D., {Salama}, F., {Allamandola}, L.~J.,
  {Martin}, P.~G., {Snow}, T.~P., {Whittet}, D.~C.~B., {Witt}, A.~N., and
  {Wolff}, M.~J., \emph{ApJ}, \textbf{592}, 947 (2003{\natexlab{b}}).

\bibitem[{Whittet} et~al.(2004)]{Whittet04}
{Whittet}, D.~C.~B., {Shenoy}, S.~S., {Clayton}, G.~C., and {Gordon}, K.~D.,
  \emph{ApJ}, \textbf{602}, 291 (2004).

\bibitem[{Nadeau} et~al.(1991)]{Nadeau91}
{Nadeau}, D., {Yee}, H.~K.~C., {Forrest}, W.~J., {Garnett}, J.~D., {Ninkov},
  Z., and {Pipher}, J.~L., \emph{ApJ}, \textbf{376}, 430 (1991).

\bibitem[{Falco} et~al.(1999)]{Falco99}
{Falco}, E.~E., {Impey}, C.~D., {Kochanek}, C.~S., {Leh{\' a}r}, J., {McLeod},
  B.~A., {Rix}, H.-W., {Keeton}, C.~R., {Mu{\~ n}oz}, J.~A., and {Peng}, C.~Y.,
  \emph{ApJ}, \textbf{523}, 617 (1999).

\bibitem[{Wucknitz} et~al.(2003)]{Wucknitz03}
{Wucknitz}, O., {Wisotzki}, L., {Lopez}, S., and {Gregg}, M.~D., \emph{A\&A},
  \textbf{405}, 445 (2003).

\bibitem[{McGough} et~al.(2005)]{Mcgough05}
{McGough}, C., {Clayton}, G.~C., {Gordon}, K.~D., and {Wolff}, M.~J.,
  \emph{ApJ, in press} (2005).

\bibitem[{Malhotra}(1997)]{Malhotra97}
{Malhotra}, S., \emph{ApJL}, \textbf{488}, L101 (1997).

\bibitem[{Wang} et~al.(2004)]{Wang04}
{Wang}, J., {Hall}, P.~B., {Ge}, J., {Li}, A., and {Schneider}, D.~P.,
  \emph{ApJ}, \textbf{609}, 589 (2004).

\end{thebibliography}

\IfFileExists{\jobname.bbl}{}
 {\typeout{}
  \typeout{******************************************}
  \typeout{** Please run "bibtex \jobname" to optain}
  \typeout{** the bibliography and then re-run LaTeX}
  \typeout{** twice to fix the references!}
  \typeout{******************************************}
  \typeout{}
 }

\end{document}